# Current-generating 'double layer shoe' with a porous sole

Anatoly B. Kolomeisky* and Alexei A. Kornyshev**

* Center for Theoretical Biological Physics, Department of Chemistry and Department of Chemical & Biomolecular Engineering, Rice University, Houston, TX 77005-1892, USA        tolya@rice.edu

** Department of Chemistry, Faculty of Natural Sciences, Imperial College London (University of Science, Technology and Medicine), London, SW7 2AZ London, UK        a.kornyshev@imperial.ac.uk

April 23, 2016

**Abstract**

**We present a principle and a simple theory of a novel *reverse electroactuator*, in which the electrical current is generated by pumping of an electrolytic liquid into nonwetting pores of a polarized electrode. The theory establishes the relationship between the variation of external pressure and the electrical current. The effective current density is amplified by high porosity of the electrode. The suggested principle can be implemented into a design of a shoe which will be generating an AC current simply by walking. Estimates of typical parameters and operation regimes of such device suggest that one can easily generate a peak current density of ~17 mA/cm$^{2.}$ This would produce some 1.7 A from each shoe at 0.65 W average power density, without hampering walking.**

## 1. Introduction

Harvesting electrical current from mechanical motion has a long history [1,2]. In everyone's car there is an electromagnetic current generator which continuously charges its battery, when car moves. Diesel motors are used for autonomic, out-of-grid current supply. Similar principles operate in wind power [3]. However, in the recent age of mobile gadgets and sustainable energy production, there is a strong need for generation of low currents, which could utilize any kind of physical motion of individuals (walking, jogging, running, even moving hands). Such generation is often called reverse electroactuation [4]. Whereas the direct electroactuation, normally used in robotics (artificial muscles) for generation of motion resulting from variation of applied voltage [5], the reverse electroactuation converts an applied force induced mechanical motion into AC voltage or current. This can be used for feeding mobile phones or tablets [1], or, alternatively for sensing purposes such as measuring pressure fluctuations in turbulent flows in hydromechanics [1], aeronautics [1], or ultrasensitive tactile sensing [6].

A principle of a new type of reverse electrowetting harvester of electrical energy has been suggested and elaborated recently by Krupenkin [7,8,27,9]. Whereas direct electrowetting changes the form of a wetting or nonwetting droplet on electrode with variation of applied voltage [10,11,12], in the reverse electrowetting the shape of a droplet is changed mechanically and this causes creation of additional voltage or generation of a transient electrical current. In Krupenkin's device [8], a liquid droplet (or many droplets) with dissolved electrolyte are compressed between electrodes or at an electrode. Compression changes the contact area between electrode and the droplet, modifying the electrical capacitance of the system. Under a constant bias voltage, a change of capacitance leads to the flows of the electronic charge and the charge of counterions from the two sides of the interface, which creates transient electrical currents. This principle is implemented into a model shoe generating alternating current from walking. Ideas along the same lines have been suggested for manipulation with ionically nonconductive, purely dielectric droplets [13], but due to much smaller capacitances in such systems (smaller than the electrical double layer capacitance), the resulting effects were much smaller than in the Krupenkin's systems.

In this paper we present a basic theory of a different kind of reverse electroactuator, although also based on the variation of the contact area of an electrical double layer with the electrode. Its principle has a several new features. The simple theory presented below relates the generated current with the variation of the applied pressure, and it is described in the context of a gadget that can be called "electrical double-layer shoe with a porous sole".

## 2. The principle of the reverse double layer actuator

The inverse electroactuator that we will discuss here will generate an AC current, caused by variation of external pressure. It can be used as for electrical energy harvesting or as a sensor of even smaller variations of pressure (in particular, as a sound amplifying device or detector of turbulence). We concentrate below on harvesting the energy from damped steps of a walker, converting that usually wasted energy into electrical current. The principle of such device is based on charging electrical double layer when electrolytic solution is being pumped into a solvophobic pore of a conductive material which is electrically polarized relative to the bulk of the electrolyte. Without an applied pressure such pore will resist penetration of the liquid into it. It will stay empty until a sufficient pressure is applied. The invasion of the electrolytic liquid into a pore of a polarized electrode will cause charge separation: counterions will tend to prevail inside the pore, co-ions not entering it – thus forming charged electrical double layers at the pore walls. The capacitance of such double layers is proportional to the contact area between the liquid and the pore. The larger the contact area, the larger the capacitance and the charge of electrolyte that the system would like to accommodate there. Since the sole of a walker may contain millions of such pores, the wall of which belonging to the same electrode, the current collected by one pore could be strongly 'amplified'.

The current generation in such system is straightforward. If the electrode is polarized negatively, the double layer will be positively charged, and while forming it will be met by the influx of negative charge of electrons to the electrode surface. If the electrode is polarized positively the forming net charge in the double layer will be negative and it will be met by the positive charge due to the depletion of electrons in the electrode. Each of these situations will cause electrical current in the network of either sign.

Importantly the electrode polarization in this system is kept fixed – it does not need to be changed, and the energy of the battery maintaining the polarization (voltage) will not be used. The current will be 'provided' only by the changing pressure, which will be pushing the electrolyte into the pore. When the pressure is off, the liquid will recede and leave the pore, and the extra charge of the electrode will flow back into the battery. Thus the current will flow in the opposite direction to what it was when the liquid was invading the pore.

A cartoon of such current generating machine is sketched in Fig.1. The picture illustrates the idea of this mode of conversion of the applied pressure into electrical current. Namely:

- ➢ The 'working electrode' (6) is porous with solvo/iono-phobic walls (hydrophobic in the case of the use of aqueous electrolyte) of micron or submicron size capillaries. For simplicity we call it microporous, although the pore sizes could be even ten nanometers small, as long as the pore radius is much greater than the thickness of the electrical double layer at the walls. The electrode is stationed at a constant disposition relative to the counter-electrode. The working electrode is polarized relative to the counter electrode (4), and the dark blue lines depict the electrical double layer regions rich in counterions, forming subject to the electrode polarization; red lines mark the double layer rich in coins forming at the counter electrode. The thickness of the double layer is of the order of 1 nm, so it is at least 100 times smaller than the radius of the pores. The counter electrode is easily wetted by the liquid.

- ➢ When the external mechanical pressure is not applied, the liquid does not fill the pores of the working electrode. With a sufficiently large applied pressure, the pores of the working electrode get impregnated with the liquid, electrical double layers form at the electrode/electrolyte interface providing a larger double layer capacitance, which itself scales with the contact area. These double layers come charged automatically as the electrode potential near the pore entrance 'filter' counterions from co-ions. (If those do not come convectively, the extra counterions will migrate from the bulk, whereas coins will leave the pore to move to the counterelectrode. The latter will cause some delay in charging, but generally the process will be dominated by convection).

- ➢ Once the capacitance is increased at the bias voltage (and thereby electrode potential) kept constant, in the case of the positive polarization of the working electrode, the double layer about it will be preferentially charged with anions. This will be responded by the same amount of negative charge of electrons leaving the electrode. For negative electrode polarization this will be cations, met by the new electrons coming to the electrode. In both cases the electrical current will be generated, positive or negative, respectively. When the pressure is off, the liquid leaves the pores of the working electrode, which will cause the reduction of capacitance, and the relaxation current will run in opposite direction.

- ➢ Although the battery is used to maintain the bias voltage, the energy of the battery is not used; apart from internal losses – the process would not discharge the battery. All the work is done by the varying applied pressure.
- ➢ Unimpeded filling of the pores with the liquid implies the measures for free removal of air from the pores, otherwise the compression of air in the pores would lead to unfavorable consequences, such as bubble formation. In other words the sole should be able to breeze. This can be done using semipermeable membranes shown at the bottom of the working electrodes (7) followed by the gas removal channel (8). Windows in the counter electrode are shown wider than the pores to imply avoidance of any resistance for the liquid flow through them. The isolating membrane (5) is also assumed not to impede the liquids flow, so that the main impedance is coming from the capillaries in the working electrode.

The cartoon is shown not in scale, real structure of the device being a matter of detailed engineering.

The task of the theory is to describe the operation of this energy harvester, i.e. to relate the varying-in-time pressure $P(t)$ with the penetration length $L(t)$, and then calculate the current $J(t)$ which such penetration will have time to cause in each cycle. The ultimate goal of the theory will be the calculation of $J(t)$ as a function of $P(t)$, for different parameters of the system. Applied voltage can be controlled independently of all other system parameters except for it may affect the double layer capacitance.

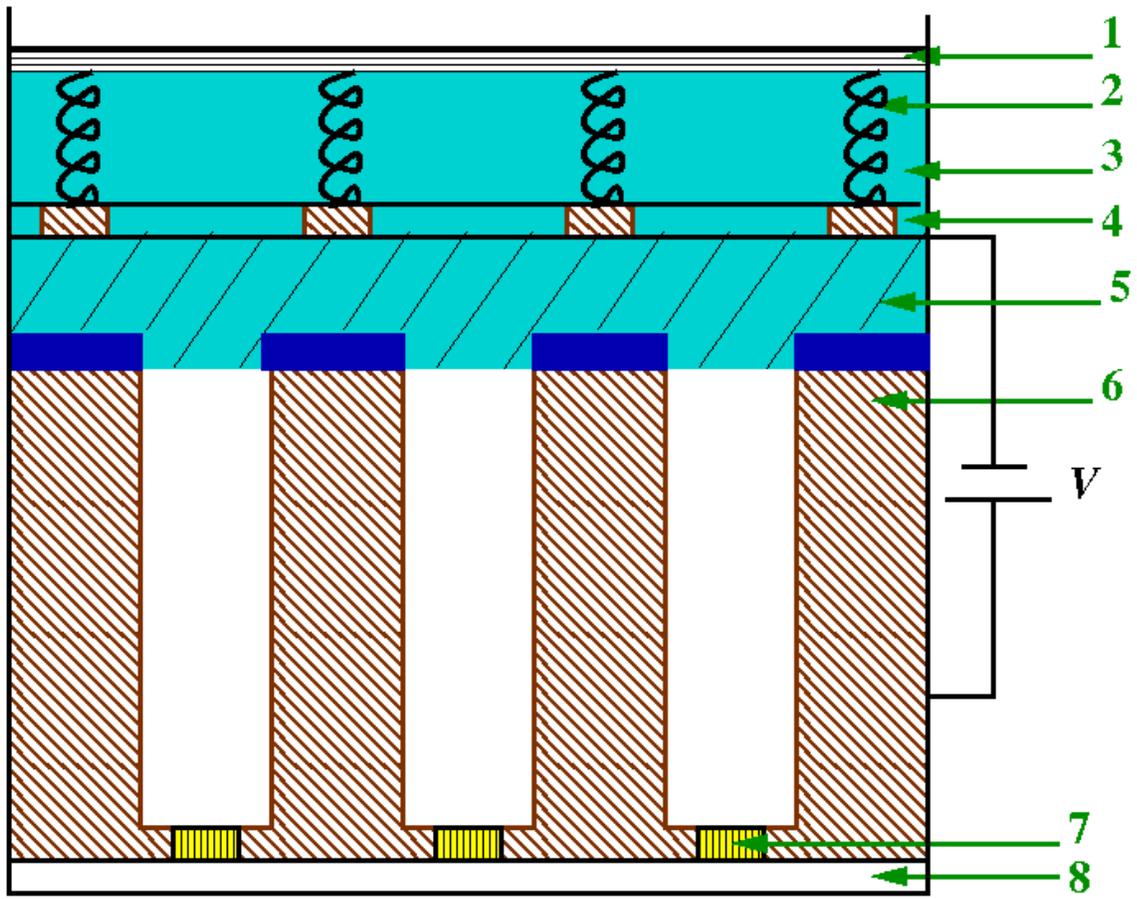

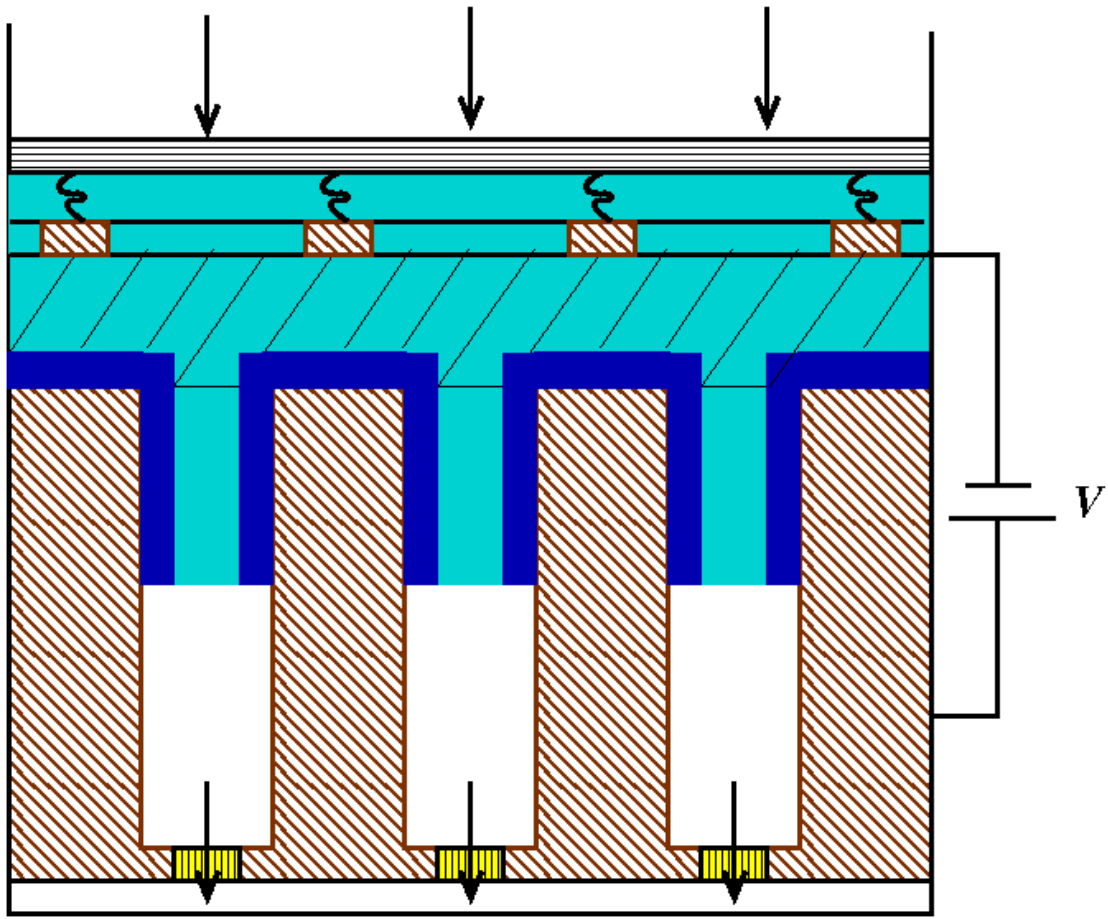

**Fig.1.** *A sketch of current generating porous sole.* Different parts are: (1) Pressure solid plate; (2) Pressure balancing spring; (3) Electrolytic solution (or ionic liquid); (4) Counter electrode with open 'windows' for the liquid flow (e.g. of muffin-tin type); (5) Liquid-permeable membrane, warranting isolation of the counter electrode from the 'working' electrode; (6) Working electrode; (7) Gas-permeable membrane; and (8) Gas removal channel. *V* denotes the bias voltage between the counter and working electrodes. The details of the device operation are explained in the text.

3. **Basic theory and results for the generated current and power**

3.1. *Dynamics of pore fillings under varying pressure*

The dynamics of pore filling under applied pressure is an independent problem of rheology. It has been studied and solved in various approximations [14,15,16]. To demonstrate the principle, we will employ here the simplest one based on the solution of Washburn equation [14,16,17]. Let us consider one single, cylindrical pore of length $L_{max}$ and radius $r$, the degree of filling of which with the electrolyte depends on the external pressure. Under the time-dependent pressure $P(t)$, the Washburn equation on the length of penetration of the liquid into the pore, $L(t)$, reads

$$L\frac{dL}{dt} = AP(t) \tag{1}$$

with the parameter $A$ defined as

$$A = \frac{r^4 + 4kr^3}{8r^2\eta}. \tag{2}$$

where $k$ is a slip coefficient and $\eta$ the viscosity of the liquid inside the pore. The slip coefficient is close to zero for wetting liquids, but for the nonwetting ones –the case we are interested in – its value may be different [9]; it has a dimensionality of distance. In nanopores, the viscosity of the electrolytic solution can generally be different from the bulk one. We will not be going to such extremes in terms of pore sizes, dealing exclusively with micropores, in order to avoid getting dynamics too slow and hysteretic, as well as not to complicate the theory. Therefore, bulk values could be safely used.

The total time dependent pressure in the pore is a sum of several main items,

$$P(t) = P_A + P_H + P_{EW} + P_{ext}(t) + P_c, \tag{3}$$

where $P_A$ stands for the atmospheric pressure, $P_H$ is a hydrostatic pressure of the electrolyte above the pores (which we assume to be constant), $P_{EW}$ is the pressure due to electrowetting, i.e., because of the effective reduction of the surface tension under the constant voltage [13], $P_{ext}(t)$ is the externally applied pressure that can be controlled, and $P_c$ is the capillary pressure. The latter may be estimated using the Laplace equation

$$P_c = \frac{2\gamma}{r}\cos\varphi, \tag{4}$$

in which $\gamma$ is a surface tension, and $\varphi$ is the contact angle.

We assume that before the application of the external force the liquid does not go into the pore, i.e., $P_c < 0$, and the total pressure is also negative,

$$P_A + P_H + P_{EW} - |P_c| < 0. \tag{5}$$

Let us also neglect $P_A$, $P_H$ and $P_{EW}$ as a first approximation. Below we will evaluate the electrowetting contribution to the pressure to justify our assumptions. It is important to note that as long as the total pressure in the pore is negative and there is no liquid in the pore, there will be no flow to the pore.

If we assume that the external pressure changes periodically with a frequency $\omega$ as

$$P_{ext}(t) = P_0(1 - \cos\omega t), \tag{6}$$

so that $P_{ext}(t=0)=0$, then Eq. (1) can be rewritten as

$$L\frac{dL}{dt} = -A[\Delta P + P_0 \cos \omega t], \tag{7}$$

where

$$\Delta P = |P_c| - P_0. \tag{8}$$

The flow into the pore is possible only when the total pressure is positive; equally it must be negative to provide the liquid flow out of the pore. Thus, generally one may be interested in a specific regime of pressure variation, and analyze Eq. (1) correspondingly. In this regime

$$|P_c|/2 \leq P_0 \leq |P_c|, \tag{9}$$

which warrants the existence of the equal intervals of time when the total pressure is positive, and those when it becomes negative, so that the liquid will be getting in as out, as needed for our device. As generally $P_0$, the amplitude of variation of external mechanic pressure, can be very large, the system can be tuned to this regime by balancing that excessive pressure with springs.

However, the optimal regime will be the particular case of Eq.(9), when $|P_c|=P_0$, i.e. $\Delta P=0$. For this case the integration of Eq.(1) simply gives

$$L(t) = \sqrt{\frac{2AP_0}{\omega}}\sqrt{(1 - \sin \omega t)} \tag{10}$$

for $t > \pi/2\omega$. This equation will of course be valid only as long as $L(t) < L_{\max} = \sqrt{\frac{2AP_0}{\omega}}$ (the total pore length). We will stay within this constraint; otherwise the sole will be gradually destroyed, as the only way to release the external stress would be to expand the pore width.

3.2. *Generated current*

Knowing the value of the total charge $Q(t)$ accumulated by the time *t* in the pore, one can calculate the current generated through one pore

$$J(t) = \frac{dQ(t)}{dt}, \tag{11}$$

The problem is then reduced to the calculation of the time dependence of the total charge $Q(t)$.

To get a feeling about $Q(t)$, we will use the fully adiabatic approximation. Namely, we assume that the ions in the liquid move very fast, momentarily charging or discharging the double layers following the change in the position of the liquid or that they convectively flow with the liquid in and out, filtered by the bias voltage. This does not mean, of course, that only counterions enter the pore, the whole double layer with its counterions and co-ions gets 'sucked-in', already polarized. This is a plausible approximation, the simplest one to start with. In such simple case, the accumulated charge in the pore is given by,

$$Q(t) = 2\pi r L(t) C \bar{U}. \tag{12}$$

Here, $C$ is the double layer capacitance per unit surface area of the pore (for the time being we consider it to be voltage independent - extension to a more general case is possible and will be considered in the follow-up publications), and $\bar{U}$ is the potential drop between the electrode and the bulk of the solution.

Substitution of Eq. (12) into Eq.(11) gives

$$J = 2\pi r C \bar{U} \frac{dL}{dt}, \tag{13}$$

In the regime of pressure variation described by Eq.(10),

$$\frac{dL}{dt} = -\sqrt{\frac{1}{2} A P_0 \omega} \frac{\cos(\omega t)}{\sqrt{(1-\sin \omega t)}} \tag{14}$$

so that the current generated from filling a single pore is given by $J = J^* \frac{\cos(wt)}{\sqrt{(1-\sin(wt))}}$ with $J^* = 2\pi r C \bar{U} \sqrt{\frac{1}{2} A P_0 \omega}$. The total current from the whole sole can be estimated as $\frac{\varepsilon S}{\pi r^2} J$ where the factor $\frac{\varepsilon S}{\pi r^2}$ is an estimated number of pores that can be accommodate on the sole of surface area $S$ and $\varepsilon$ is porosity. Thus the characteristic current per unit surface area of the sole, i.e. thus defined 'current density' is given by $\frac{\varepsilon}{\pi r^2} J$, resulting in

$$j = -j^* \frac{\cos(\omega t)}{\sqrt{(1-\sin \omega t)}}, \quad j^* = \frac{2\varepsilon}{r} C \bar{U} \sqrt{\frac{1}{2} A P_0 \omega} \tag{15}$$

where $j^*$ can be called a *characteristic current density.*

The latter is neither the maximum current density, nor the average current density over a quarter of a period. Indeed, $\left|\frac{\cos(\omega t)}{\sqrt{(1-\sin \omega t)}}\right| = \sqrt{1+\sin(\omega t)}$; thus, the maximal value of this function is $\sqrt{2} = 1.41$. The value averaged over a quarter of a period $\frac{2\omega}{\pi} \int_{\pi/2\omega}^{\pi/\omega} dt \frac{\cos(\omega t)}{\sqrt{(1-\sin \omega t)}} = \frac{4}{\pi} = 1.27$. But the characteristic current density gives the scale of the effect. It is thus worth to estimate it. For $C = 0.1$ F/m$^2$, $\bar{U}$=0.3 V, $r=10^{-6}$ m, $A=10^{-10}$ m$^2$/(Pa s), $P_0 =10^5$ Pa, and 1s period of pressure oscillation, i.e. $\omega = 2\pi$ s$^{-1}$, gives 17 mA/cm$^2$. The whole sole of 100 cm$^2$ will thus give 1.7 A and two soles (two feet) will give 3.4 A.

Figure 2 shows the resulting time dependence of the current for this case (by convention, application starts from the quarter of a period).

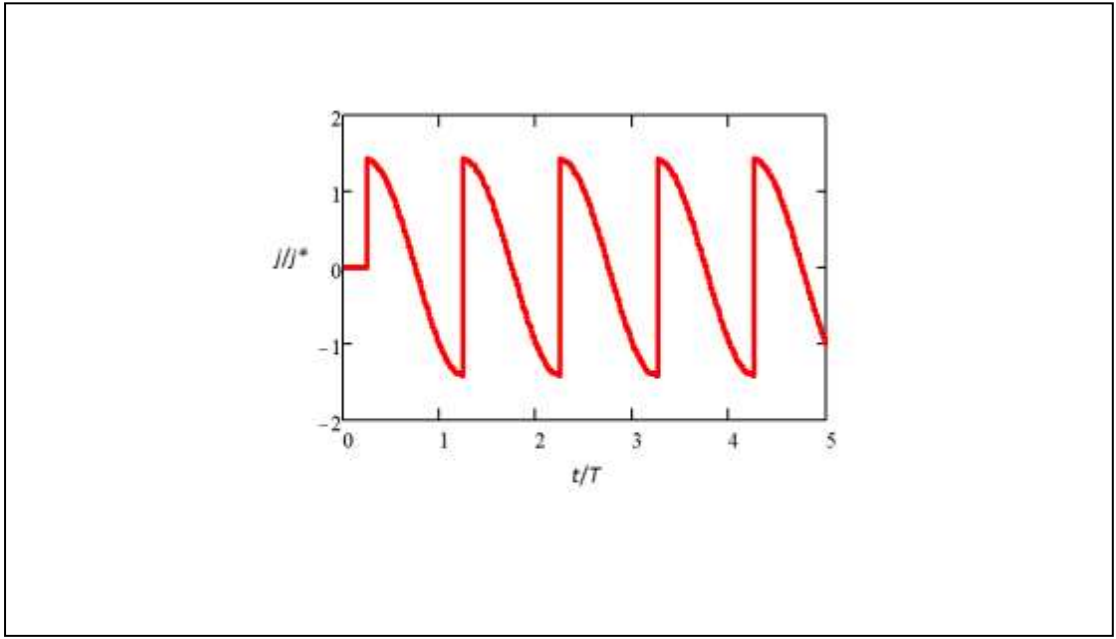

**Fig.2.** *The current generation in adiabatic approximation.* The reduced current is shown for 5 periods (each - $T = 2\pi/\omega$) of oscillating pressure. Note that if the period is two times shorter (running instead of walking) the characteristic current $j^*$, will be the $\sqrt{2} = 1.41$ times larger. The faster you run, the (square root times) more current you generate.

One should note here that for this choice of parameters the maximal pore length $L_{max}$ is less than 1mm, which is a reasonable value for a shoe. Also, the pressure due to electrowetting can be estimated as [18]

$$P_{EW} = \frac{c\bar{U}^2}{r}. \tag{16}$$

Then using the same parameters we estimated it as $P_{EW} = 10^4$ Pa, which is only 10% of $P_0$, justifying our approximation that neglected this contribution in the first approximation.

3.3. *Power density*

We can calculate the generated average power density per unit surface area of a sole, as $W=<J>U$, averaging the current over the quarter of the period. With $J$ given by equation (15), we obtain a simple formula

$$W = 1.27\, j^{*}\bar{U} = \frac{2.54\varepsilon}{r} C\bar{U}^{2} \sqrt{\frac{1}{2} AP_{0}\omega} \ . \tag{17}$$

For the same set of parameters that we used above to estimate the characteristic current density, we get for the power generated per 1 cm$^2$ of a sole 1.27 x 0.017 A x 0.3 V=6.5 mW. The power generated from two soles can thus be estimated as 1.3 W.

This figure is 1000 times greater than those reported for already perfectly working prototypes of piezoelectric cymbal harvesters of electrical energy from walking [19], and comparable to the power output proposed by Krupenkin's device. [7,8]

Krupenkin's device offers, 1 W per shoe, by suggesting a beguiling idea of patched electrodes. Indeed, if the surface over which a droplet spreads upon compression, contains alternating dielectric and metallic patches, then the moment the droplet spreads over one metallic patch but yet has not reached another one, the capacitance stops increasing, the separated charge stalls, and the current drops to zero. However after the droplet spreads wider and reaches the next metallic patch it will again increase the capacitance, new charge separation will take place, and the current will peak. Building a system of micropatches, Krupenkin, was able to generate many spikes of the current over one step, claiming that this will amplify the generated power. The same type of patching can be implemented in our system, but the investigation of conditions when such architecture will be amplifying the power requires special investigation (which we will be done in the next work). Obviously, when the metallic patches are too rare, the power will be lower than in a continuous metallic electrode system.

It is also important to compare the generated electrical power with the mechanical power produced by humans when walking. Experiments indicate that this power is of order of 100-500 W [20], which is still larger than can be harvested by the currently suggested devices including this one. There is a lot of reserves to be explored, but a good news here that walking with the AC current-generating double-layer shoe will probably not be a problem due to resistance of the sole, i.e., it will not be difficult to push the electrolyte liquid into the pore.

3.4 *Critical evaluation of the made assumptions*

The process of generating the electrical current in the double layer shoe is generally a complex problem where hydrodynamics is coupled with wetting, electrowetting, and kinetic phenomena. Therefore, it is important to critically evaluate the presented theoretical arguments and the effects that we left behind this first simplistic consideration.

The first question one might ask is about the applicability of the simple Washburn equation for the description of the dynamics of pore filling by the ionic liquid. Strictly speaking, this equation assumes a steady-state flow into the pore, where the inertia of fluid is neglected in comparison with the viscous friction [11]. This also means that the current follows the flow without inertia. We can estimate the relaxation time into such stationary behavior [11],

$$\tau_c \cong \frac{\rho r^2}{4\eta}, \tag{18}$$

where $\rho$ is a density of the fluid, $\eta$ is a viscosity of the fluid and $r$ is the size of the pore. Using our standard set of parameters and assuming that the density and viscosity of the liquid are of the same order as corresponding values for water, we obtain that this time is less than a microsecond. But this is much smaller than the typical times scales (seconds) considered in our model. Thus, this estimate justifies the application of Washburn equation for our calculations.

Another important question is that our approach assumes a laminar flow, while for turbulent flow the analysis is obviously not correct. To evaluate if the flow in our system is laminar or turbulent we need to estimate Reynolds number. For our system, it is given by

$$Re = \frac{\rho \frac{dL}{dt} L}{\eta}. \tag{19}$$

Again using our standard set of parameter values, it can be calculated that $Re \approx 2$, which is much smaller than the onset of the turbulence ($Re \approx 2000$), and hence the assumption of a laminar flow is fully justified.

As mentioned in the beginning of the paper, the thickness of the double layers at the pore wall was assumed to be much smaller than the pore radii, but in any case to make the capacitance larger we need high concentrated electrolytes, so as long as the pore size is above 10 nm, that assumption would be guaranteed.

We also neglect the effect of electroosmotic flows, because by the assumption of the model the incoming liquid is already fed by the proper amount of counterions. Generally one should consider electrokinetic phenomena together with convective flows, but this will be investigated with a more advance theory. A summary of the main avenues for the future development of the theory of the current generation in the 'double layer shoe' is given in the next section.

## 4. Concluding remarks

We have suggested a scenario of generation of the electrical current by walking different from the suggested earlier, having developed a simple, perhaps oversimplified theory of the liquid electrolyte filling hydrophobic pores under the varying pressure at static bias voltage. That theory utilizes (i) the Washburn law of pore filling under pressure, and (ii) adiabatic response of ions tending to recharge the pore in the course of the formation of electrical double layers or instantaneously flowing in and out of the pore with the liquid, 'filtered' by the static polarization of the electrode.

This scenario will benefit from stronger hydrophobicity of the surface. Indeed, then higher voltages could be applied without a risk of electrowetting of the interior surfaces of the pores, which itself may result in spontaneous pore filling with the liquid, without any external force.

There is a limit on applied voltages, however, as we do not want onset of any electrochemical reactions at the interface. In other words, the voltage should lie within the 'electrochemical window', which is different for every electrode/electrolyte combination [21]; it can be made larger for organic electrolytes in organic solvent and in particular in ionic liquids [22] (also suggested to be used by Krupenkin). However, changing the liquid phase, one must check in which direction will be the changes of the ability to wet the used electrode material.

For very narrow (<100 nm) close ended pores, the electrode surface may become superhydrophobic even if it is not particularly hydrophobic in its flat, nonporous state [23,24]. Close pockets of air will further prevent spontaneous penetration of the liquid inside the pores. Such 'ultrahydrophobicity'[25] however will not be straightforward to exploit. Indeed, it will lead to new effects under external pressure, such as transitions from Cassie-Baxter to Wenzel states [23,24,25], which would change the laws of liquid propagation inside the pore, and thus should be a subject of separate investigation.

It might be tempting to use solvent-free ionic liquids as electrolytes because of their nonvolatile nature. However two obstacles may be encountered here. The electrode must be made ionophobic, but although ionopobicity was predicted to be benefical for charging dynamics of supercapacitors with ultrananostructured electrodes [26,27], this property has not been yet successfully experimentally implemented. Pure ionic liquids may appear to be too viscous, and this, according to Eqs. (2) and (17), will reduce the generated power. To diminish viscosity, to 'lubricate' ionic liquids, small addition of organic solvent is often used [28], but the issue of the overall phobicity of the electrode to such solution remains to be fulfilled for our gadget to function.

A more detailed theory of dynamics of pore filling should go beyond the simplistic Washburn-like theory considering more sophisticated rheological scenarios of pore filling/de-filling under pressure, potentially including capillary hysteresis. It should describe explicitly dissipation and generation of heat, and invoke a more detailed account of the counterion and coion motion and, generally, of electrokinetic phenomena [29,30]. The latter will lead to more complicated equations that will be considered in a future work. Furthermore, in a practical device one must consider the effects of more complicated porous structures in the sole (tortuosity of the pores, pore radius and pore length distribution) and their deformability under the varying pressure.

All these effects should be studied in the future detailed theory of the current generation based on the above described principle, which opens a whole new area of research. They indeed, need to be investigated because all of them will lead to losses of power. Thus the estimates presented above refer to maximal possible power. However, before starting to develop the theory in the mentioned directions, the first priority will be experimental verification of the predictions of this simplest theory. Naturally, before going into the complexity of porous electrodes, the experiments should be done in a well-defined single microfluidic pore. Such experiments will test the suggested principle of the current generation. The mentioned developments will be needed to explore all technical aspects of the *current-generating double layer shoe with a porous sole*, and whether in practical terms it could compete with piezoelectric harvesters or may appear to have some benefits as compared to Krupenkin's.

In conclusion, we note that, as suggested to us by M.Bazant, another version of this device is possible using two immiscible liquids, one without electrolyte and strongly wetting the electrode surface, and the other one containing electrolyte and not wetting the electrode. The theory will not change in this case, just the value of some parameters will, such as *A* and γ. But replacement of air in the channels with the liquid might cause difficulties related to a capillary ('Laplace') instability [31], which under certain conditions could cause viscous fingering[32]. It may furthermore cause complications related with removal of non-electrolytic liquid from the channel upon invasion of the electrolytic one. This extension may, however, be considered in future work.

Last but not least, the device architecture may be modified, by turning the structure upside down, so that the electrolytic solution (or ionic liquid) will be underneath, invading the pores when the porous electrode moves down on it. If in the basic equations the gravity of the liquid is neglected, the equations will essentially be the same as described and used above. Will such structure be more practical, is again a matter of engineering.

It should be also mentioned that there are other devices that exploit various concepts of energy harvesting [33,34,35,36]. They utilize salinity differences [33], electrolyte concentration gradients [34], mixing entropy in channels [35] and mechanical modulations of water bridges between conducting planes [36] as a source of energy production. Similarly to our device, all of them also use the electrical double layer capacitors. However, the power production is much smaller than predicted in our system. It will be important to explore the connections between different concepts of energy harvesting with double-layer capacitors in order to develop more efficient devices.

**Acknowledgements**

The authors are thankful to Martin Bazant (MIT), Douglas Natelson and Peter Wolynes (Rice University), and Eugene Kolomeisky (University of Virginia) for reading the manuscript and for most useful critical comments. AAK thanks also discussions with Joshua Edel, Lazar Obradovich and Gunnar Pruessner (Imperial College), and Michael Urbakh (University of Tel Aviv). ABK acknowledges the support from the NSF (Grant CHE-1360979) and from the NSF-sponsored (Grant PHY-1427654) Center for Theoretical Biological Physics at Rice University.